\documentclass{emulateapj}

\usepackage{graphicx}
\usepackage{psfig}
\usepackage{apjfonts}

\def\cm{\textrm{cm}}
\def\Br{\textrm{Br}}
\def\gram{\textrm{g}}
\def\sec{\textrm{s}}
\def\km{\textrm{km}}

\def\kpc{\textrm{kpc}}
\def\pc{\textrm{pc}}
\def\Mpc{\textrm{Mpc}}

\def\gcm2{\textrm{g}~\textrm{cm}^{-2}}
\def\ergscm3{\textrm{erg}~\textrm{s}^{-1}~\textrm{cm}^{-3}}
\def\ergcm3{\textrm{erg}~\textrm{cm}^{-3}}
\def\gscm2{\textrm{g}~\textrm{s}^{-1}~\textrm{cm}^{-2}}
\def\ergcmK34{\textrm{erg}~\textrm{cm}^{-3}~\textrm{K}^{-4}}
\def\cms31{\textrm{cm}^{-3}~\textrm{s}^{-1}}
\def\cmg21{\textrm{cm}^{2}~\textrm{g}^{-1}}

\def\phFluxUnits{\textrm{cm}^{-2}~\textrm{s}^{-1}}

\def\GeV{\textrm{GeV}}
\def\TeV{\textrm{TeV}}

\def\Myr{\textrm{Myr}}
\def\Gyr{\textrm{Gyr}}

\def\Msun{\textrm{M}_{\sun}}
\def\Lsun{\textrm{L}_{\sun}}
\def\AU{\textrm{AU}}
\newcommand{\mean}[1]{\ensuremath{\langle #1 \rangle}}

\begin{document}

\title{Primordial Black Holes as Dark Matter: Almost All or Almost Nothing}
\author{Brian C. Lacki\altaffilmark{1,2} \& John F. Beacom\altaffilmark{1,2,3}}
\altaffiltext{1}{Department of Astronomy, The Ohio State University, 140 West 18th Avenue, Columbus, OH 43210, lacki@astronomy.ohio-state.edu.}
\altaffiltext{2}{Center for Cosmology and AstroParticle Physics, The Ohio State University, 191 West Woodruff Avenue, Columbus, OH 43210, beacom@mps.ohio-state.edu.}
\altaffiltext{3}{Department of Physics, The Ohio State University, 191 West Woodruff Avenue, Columbus, OH 43210.}

\begin{abstract}
Primordial black holes (PBHs) are expected to accrete particle dark matter around them to form ultracompact minihalos (UCMHs), if the PBHs themselves are not most of the dark matter.  We show that if most dark matter is a thermal relic, then the inner regions of UCMHs around PBHs are highly luminous sources of annihilation products.  Flux constraints on gamma rays and neutrinos set strong abundance limits, improving previous limits by orders of magnitude.  Assuming enough particle dark matter exists to form UCMHs, we find that $\Omega_{\rm PBH} \la 10^{-4}$ (for $m_{\rm DM} c^2 \approx 100\ \GeV$) for a vast range in PBH mass.  We briefly discuss the uncertainties on our limits, including those due to the evolution of the UCMH luminosity as it annihilates.
\end{abstract}


\section{Introduction}
The early Universe was extremely smooth, but it is possible that there were rare but large perturbations.  Any large perturbations ($\delta \rho / \rho \ga 0.3$) collapsed to form Primordial Black Holes (PBHs; \citealt{Hawking71}).  PBHs are expected to acquire a halo from the surrounding particle dark matter background, if they themselves do not make up most of the dark matter \citep{Mack07,Ricotti07,Ricotti08,Ricotti09}.  These halos are called Ultracompact Minihalos (UCMHs; \citealt{Scott09}).  

The abundance of PBHs with masses above $1000\ \Msun$ is strongly constrained ($\Omega_{\rm PBH} \la 10^{-8}$), using the effects of baryonic accretion on the CMB energy spectrum \citep{Ricotti08}, and that of masses below $\sim 10^{-15} \Msun$ is constrained by the non-observation of gamma rays from PBH evaporation \citep[see][and references therein]{Carr09}.  In the planetary--stellar mass regime, there are modest constraints ($\Omega_{\rm PBH} \la 10^{-1}$) from microlensing (e.g., \citealt{Alcock98,Alcock01,Tisserand07}, but see \citealt{Alcock00}) and the dynamics of structure formation and widely separated binary stars (\citealt{Yoo04}; but see also \citealt{Quinn09}).  Over a huge range of possible PBH masses ($10^{-15} - 10^{-9}\ \Msun$), the constraints in their abundances are weak or non-existent \citep[e.g.,][]{Seto07,Abramowicz09}.  

The dark matter UCMHs around PBHs allow new ways to constrain the abundance of PBHs.  \citet{Ricotti09} suggested searching for the microlensing signature of UCMHs, also pointing out that there could be a gamma-ray signal from dark matter annihilation in the dense core of the UCMH.  \citet{Scott09} developed this further, considering the gamma-ray signal from individual, nearby UCMHs of various masses, and their detectability with \emph{Fermi}.  

We develop strong new constraints on dark matter annihilation in UCMHs around PBHs, using general assumptions that can be evaded only in exotic dark matter models.  These limits apply if PBHs do not make up all of the dark matter, so that enough particle dark matter exists to form UCMHs.  We strive for order-of-magnitude accuracy.

\section{Dark Matter Annihilation}
\label{sec:Annihilation}

We assume dark matter is a thermal relic, following annihilation freezeout in the early Universe, consisting of Weakly Interacting Massive Particles (WIMPs).  We assume it may self-annihilate in the late Universe (i.e., there is no large WIMP-antiWIMP asymmetry), and that the final states are Standard Model particles.  For thermal relic dark matter, $\Omega_{\rm DM} h^2 = 3 \times 10^{-27} \cm^3 \sec^{-1} / \mean{\sigma_A v}$, relating the dark matter density to its thermally-averaged annihilation cross section \citep{Kolb90}.  We take $\Omega_{\rm DM} = 0.3$ and $h = H_0 / (100\ \km\ \sec^{-1}\ \Mpc^{-1}) = 0.7$.  We consider masses above the GeV range, probed at colliders, and below a PeV, set by the unitarity bound.  We use the relic $\mean{\sigma_A v}$, assuming it independent of $v$; this holds for standard s-wave annihilation.  Our results cannot be evaded by reducing $\mean{\sigma_A v}$, as this would lead to an unacceptably large $\Omega_{\rm DM}$.

While $\mean{\sigma_A v}$ is fixed, the particular Standard Model final states are not; \emph{however, strong constraints on dark matter annihilation can be set in any case} \citep{Beacom07}.  We then limit the abundance of UCMHs from constraints on dark matter annihilation, and thus the abundance of PBHs themselves.  

Constraints on dark matter annihilation from gamma-ray signals are well known.  Direct annihilation produces a spectral line at $m_{\rm DM} c^2$ \citep{Mack08}.  Annihilation into quarks, charged leptons, or gauge bosons produces a gamma-ray continuum when those particles decay. Electron-positron final states produce gamma rays, through internal bremsstrahlung, with branching fraction $\Br (\gamma) \approx \alpha \approx 0.01$.  This process is an electromagnetic radiative correction; it is independent of surrounding matter density, occurs in any process involving charged particles, and produces gamma rays with energies up to $m_{\rm DM} c^2$ \citep{Beacom05,Bell09}.  Additional radiation is produced through synchrotron and Inverse Compton energy-loss processes. 

It might be thought that neutrinos are invisible annihilation products; in fact, there are stringent bounds on them \citep{Beacom07,Yuksel07}.  The sheer size of neutrino telescopes like IceCube compensates for the small detection cross section, and the atmospheric neutrino background falls off steeply with energy.  Other Standard Model annihilation products are intermediate to the above cases.  

\section{UCMH Profiles and Luminosities}
\label{sec:UCMHProfile}

According to \citet{Mack07} and \citet{Ricotti07}, a UCMH (with or without a PBH) at $z \approx z_{\rm eq}$ with dark matter mass $M_{\rm eq}$ has a truncation radius 
\begin{equation}
\label{eqn:Rtrz1000}
R_{\rm tr} (z) = 1300\ \AU\ \left(\frac{M_{\rm eq}}{\Msun}\right)^{1/3} \left(\frac{1 + z}{1 + z_{\rm eq}}\right)^{-1}.
\end{equation}
UCMHs with PBHs at matter-radiation equality have $M_{\rm eq} = M_{\rm PBH}$.  The dark matter mass density within $R_{\rm tr} (z)$ is the same \emph{no matter what the PBH mass is}, with a number density $n_{\rm tr} \approx 1.8 \times 10^5\ \cm^{-3}\ m_{100}^{-1}$ at $z_{\rm eq}$, where $m_{100} = m_{\rm DM} c^2 / (100\ \GeV)$.  Within $R_{\rm eq} = R_{\rm tr} (z_{\rm eq})$, the UCMH density profile goes as $\rho \propto r^{-3/2}$, because the PBH dominates the mass; at larger radii, a $r^{-9/4}$ density profile holds \citep{Bertschinger85}.  This density profile is shallower than in \citet{Ricotti09}, which assumed $\rho \propto r^{-9/4}$ throughout, and that adiabatic contraction increased the dark matter density further.  

WIMPs have nearly radial orbits around the PBH, sampling a wide range of densities in each orbit.  Thus, the dark matter annihilating in the inner regions of the UCMH is often stored much farther out.  To answer whether WIMPs with apocenters of $R_{\rm eq}$ survive to the present day, we calculate the number of annihilations a WIMP is expected to experience during the Universe's history.  This is the number of annihilations per orbit times the number of orbital periods in a Hubble time: 
\begin{equation}
\label{eqn:Nann}
\mean{N_{\rm ann}} = 2 \left(\frac{t_H}{P}\right) \int_{0}^{\pi} n_{\rm DM} (r) \mean{\sigma_A v} \frac{dt}{d\theta} d\theta,
\end{equation}
where we have defined $\theta = 0$ as apocenter and $P$ is the orbital period.  A WIMP can only annihilate once, so the density profile is valid at present only if $\mean{N_{\rm ann}} < 1$.  Assuming the orbit is Keplerian and nearly radial,
\begin{equation}
\mean{N_{\rm ann}} \approx \frac{3 \mean{\sigma_A v} M_{\rm PBH} t_H}{2 \pi^2 m_{\rm DM} R_{\rm eq}^{3/2} r_a^{3/2}} \left[1 + \ln \left(\frac{\pi G M_{\rm PBH}}{v_a^2 r_a}\right)\right],
\end{equation}
where $r_a$ is the apocenter radius and $v_a$ is the WIMP's tangential velocity at apocenter.\footnote{We approximated $K(k) = \int_0^{\pi / 2} d\theta / \sqrt{1 -k^2 \sin^2 \theta}$ as $1/\sqrt{-k^2} [1 + \ln(\pi\sqrt{-k^2}/2)]$, where $k^2 = 2e / (e - 1)$, and $e$ is the eccentricity.  This is valid if $k^2 \ll -1$, which holds for nearly radial orbits.}  Using Eq.~\ref{eqn:Rtrz1000} to find $R_{\rm eq}$ and 
\begin{equation}
\label{eqn:vTanDM}
v_a (r_a = R_{\rm eq}) \approx 8.1\ \cm\ \sec^{-1} \left(\frac{z}{z_{\rm eq}}\right)^{-1/2} \left(\frac{M_{\rm PBH}}{\Msun}\right)^{0.28},
\end{equation}
from the bulk velocity dispersion in \citet{Ricotti09}, we find 
\begin{equation}
\mean{N_{\rm ann}} \approx 0.03\ m_{100}^{-1} \left(\frac{t_H}{10\ \Gyr}\right) \left[1 - 0.005 \ln \left(\frac{M_{\rm PBH}}{\Msun}\right)\right].
\end{equation}
Therefore, WIMPs with apocenters of $R_{\rm eq}$ mostly survive to the present, despite venturing much further in.

The accretion and resultant WIMP orbits at $R_{\rm eq}$ must be nearly radial: $\sigma_{\rm DM} \ll \sqrt{2 G M_{\rm PBH} / R_{\rm eq}}$.  Using $v_a$ (Eq.~\ref{eqn:vTanDM}) for the initial velocity dispersion $\sigma_{\rm DM}$, we find accretion is radial for all reasonable PBH masses.  WIMPs also have residual thermal dispersion, $v_{\rm therm} \approx 1.3\ \cm\ \sec^{-1} m_{100}^{-1} (z / z_{\rm eq})^{-1}$; conservatively, we find radial accretion at $R_{\rm eq}$ holds for $M_{\rm PBH} \ga 2 \times 10^{-15} m_{100}^{-3} (z / z_{\rm eq})^{-3} \Msun$.  Accretion at later times than $z \approx 3000$, at smaller radius than $R_{\rm eq}$, or from the low end of the WIMP velocity distribution might still form a UCMH; so our limits may roughly hold for smaller PBHs.  Finally, WIMPs must miss the central PBH.  WIMPs with an apocenter $r_a$ have a pericenter $r_p \approx v_a^2 r_a^2 / (2 G M_{\rm PBH})$; if the apocenter is $R_{\rm eq}$, then
\begin{equation}
r_{\rm p,eq} \approx 6.3 \times 10^{-6} \AU \left(\frac{M_{\rm PBH}}{\Msun}\right)^{0.23}.
\end{equation}
Thus, WIMPs with apocenters of $R_{\rm eq}$ remain outside the Schwarzschild radius if $M_{\rm PBH} \la 1740\ \Msun$.  

The UCMH annihilation luminosity is $L_{\rm ann} = \int_{\rm r_{\rm min}}^{R_{\rm eq}} 2 \pi r^2 n(r)^2 \mean{\sigma_A v} m_{\rm DM} c^2 dr$, or 
\begin{equation}
L_{\rm ann} = \frac{9}{32 \pi} \frac{M_{\rm PBH}^2 c^2 \mean{\sigma_A v}}{m_{\rm DM} R_{\rm eq}^3} \ln\left(\frac{R_{\rm eq}}{r_{\rm min}}\right),
\end{equation}
where $r_{\rm min}$ is a minimum-radius cutoff.  Even if WIMPs with small apocenters annihilate away by the present, WIMPs on radial infall orbits with larger apocenters generate a time-averaged density profile of $\rho \propto r^{-3/2}$ between apocenter and pericenter.\footnote{Consider a shell of non-interacting WIMPs at apocenter $r_a$ and with identical tangential speeds.  The time-averaged mass at each radius is proportional to the time they spend there: $\mean{dM(r)} \propto dr / v_r$, where $v_r$ is the radial velocity.  For nearly radial orbits, $v_r \approx \sqrt{2GM/r}$, except near pericenter and apocenter.  Thus $\mean{dM(r)} \propto r^{1/2} dr$; since $\mean{dM(r)} = 4 \pi r^2 \mean{\rho(r)} dr$, $\mean{\rho(r)} \propto r^{-3/2}$ except near pericenter and apocenter.}  Therefore, we use $r_{\rm p,eq}$ as $r_{\rm min}$.  The UCMH luminosity is 
\begin{equation}
\label{eqn:LannPBH}
L_{\rm ann} \approx 24\ \Lsun\ \left(\frac{M_{\rm PBH}}{\Msun}\right) m_{100}^{-1},
\end{equation}
ignoring a very small logarithmic term in $M_{\rm PBH}$.  The linear scaling with $M_{\rm PBH}$ arises because the UCMH mass scales linearly with PBH mass, and the WIMP density (and annihilation lifetime) at $R_{\rm eq}$ is always the same.  When $\rho \propto r^{-3/2}$, each decade in minihalo radius produces the same annihilation luminosity; the total luminosity therefore is not strongly dependent on the WIMP orbits' eccentricities.  The eccentricities themselves depend weakly on $M_{\rm PBH}$, since $r_{\rm a, eq} / r_{\rm p, eq} \propto M_{\rm PBH}^{0.1}$.  Thus the total luminosity is understood as the mass of the UCMH halo (proportional to $M_{\rm PBH}$) annihilating over a \emph{fixed} annihilation timescale (set by $n_{\rm tr}$) multiplied by some slowly varying logarithmic factor accounting for the UCMH's inner regions.  

A steeper density profile inside $R_{\rm eq}$ would increase the luminosity but shorten the WIMP survival time.  Eq.~\ref{eqn:Nann} implies, for WIMPs with a $r^{-9/4}$ density profile and $r_a = R_{\rm eq}$, $\mean{N_{\rm ann}} = 1500 m_{100}^{-1} (M_{\rm PBH} / \Msun)^{0.08}$, so $t_{\rm ann} \approx 7 m_{100}^{-1}\ \Myr$.  For a $r^{-2}$ profile, $\mean{N_{\rm ann}} = 22 m_{100}^{-1} (M_{\rm PBH} / \Msun)^{0.05}$ ($z_{\rm ann} \approx 10$ when $m_{100} = 1$).  Efficient annihilation introduces $\sim \Omega_{\rm PBH}$ worth of high energy radiation into the Universe.  At high redshift, the CMB energy spectrum and reionization history severely constrain the gamma-ray and cosmic ray injection rate \citep[e.g.,][]{Fixsen96,Padmanabhan05,Mapelli06}.  At low enough redshift, the gamma-ray and neutrino backgrounds still constrain the annihilation products directly.  Thus, it would be difficult to have $\Omega_{\rm PBH} \approx \Omega_{\rm WIMP}$, unless perhaps the annihilation were entirely into neutrinos.  Furthermore, such efficient annihilation would mean that present UCMHs around PBHs have depleted inner halos, with consequences for microlensing searches \citep{Ricotti09}.  More work on the effects of annihilation in steep density profiles is needed.

\subsection{Assumptions and Other Considerations}
\label{sec:OtherConsiderations}

This annihilation luminosity depends on some simple considerations.  We assume that PBHs have not evaporated yet, which is valid for $M_{\rm PBH} \ga 10^{-18}\ \Msun$; and that most of the dark matter is made of WIMPs and not PBHs, otherwise there would be too few WIMPs to form a UCMH.

We require that the density profile along a WIMP's orbit not evolve significantly.  Most importantly, we require that dark matter accretion after $z_{\rm eq}$ does not increase the mass within $R_{\rm eq}$.  Such accretion would initially increase the UCMH luminosity, but would also shorten the time WIMPs survive in the UCMH.  We also assume that annihilation of WIMPs with small apocenters does not affect our calculations.  As a limiting case, if WIMPs annihilate into a constant density core with $\rho = \rho(R_{\rm eq})$, then $L_{\rm ann} \approx 0.42 \Lsun (M_{\rm PBH} / \Msun) m_{100}^{-1}$; this weakens our bounds by a factor of $\sim 60$, but still implies $\Omega_{\rm PBH} \ll 0.1$ for $m_{\rm DM} c^2 \la 10\ \TeV$.  Furthermore, since the PBH dominates the mass within $R_{\rm eq}$, adiabatic contraction should be unimportant.

Perhaps the most important other effects we do not consider are those of accreted baryonic material, which may cool radiatively and collapse efficiently.  Adiabatic contraction will not be important if the PBH dominates the mass within $R_{\rm eq}$.  However, baryonic matter can be optically thick to gamma rays, reducing the apparent UCMH luminosity.  We estimate the optical depth as $\tau \approx n_e \sigma_T R_{\rm eq}$.  If the mass of baryons within $R_{\rm eq}$ is $f_b M_{\rm DM} \approx f_b M_{\rm PBH}$, then $\tau \approx 0.5 f_b (M_{\rm PBH} / \Msun)^{1/3}$.  Klein-Nishina effects will reduce this optical depth.   Thus for smaller PBHs ($M \la 10\ \Msun$), we can ignore baryonic opacity.  Neutrino limits are unaffected by opacity.

\section{Cosmic Gamma-ray Background Constraints}
\label{sec:GRBackground}
Annihilation of dark matter can produce gamma rays with significant power near $m_{\rm DM} c^2$.  Gamma rays below $\sim 100\ \GeV$ contribute directly to the extragalactic background.  Gamma rays with energy $\ga 100\ \GeV$ for $z \ga 1$ cascade down in energy by pair production and Inverse Compton processes with ambient photons, contributing to the background at lower energies.

The gamma-ray emissivity of the Universe at energies above 100 MeV is limited by EGRET observations to $Q_{\rm max} = 8 \times 10^{-35}\ \ergscm3$ \citep{Coppi97}.  The number density of PBHs is then limited to be $n_{\rm PBH} \la Q_{\rm max} / (L_{\rm ann} \Br(\gamma))$, where $\Br (\gamma)$ is the branching fraction into gamma rays.  If dark matter annihilates into charged particles, there must be internal bremsstrahlung, with branching ratio $\Br (\gamma) \approx \alpha \approx 0.01$, which we take as a minimum branching fraction.  This number density is converted into a limit on $\Omega_{\rm PBH}$ by multiplying by the mass of the PBH and dividing by the critical density, $\rho_c = 9.20\ h_{70} \times 10^{-30}\ \gram\ \cm^{-3}$, so that $\Omega_{\rm PBH} \la Q_{\rm max} M_{\rm PBH} / (\rho_c L_{\rm ann} \Br (\gamma))$:
\begin{equation}
\label{eqn:CosmicGammaLimits}
\Omega_{\rm PBH} \la 1.9 \times 10^{-5}\ m_{100} \left(\frac{\Br (\gamma)}{0.01}\right)^{-1}.
\end{equation}
The upper limits on $\Omega_{\rm PBH}$ are essentially independent of PBH mass: the number of PBHs for a given $\Omega_{\rm PBH}$ scales as $M_{\rm PBH}^{-1}$, but the luminosity of each scales as $M_{\rm PBH}$.

\section{Milky Way Gamma-Ray Constraints}
\label{sec:MilkyWayConstraints}

Stronger constraints can be obtained by taking advantage of the higher-than-average number density of PBHs in the Milky Way.  PBHs in the Milky Way are close enough that annihilation gamma rays are not attenuated; such gamma rays do not have to compete with the entire gamma-ray background above 100 MeV, but only with that near $m_{\rm DM} c^2$.  

Suppose the density of PBHs, $n_{\rm PBH} (\vec{s})$, tracks the dark matter density.  Then the integrated gamma-ray intensity on a line of sight out of the Milky Way is $I = \frac{1}{4\pi} L_{\rm ann} \mean{n_{\rm PBH}} \Br (\gamma) \int \delta (\vec{s}) ds$, where $\delta$ is the dark matter overdensity over the average cosmic dark matter density and $\mean{n_{\rm PBH}}$ is the average PBH density in the Universe.  Then the abundance of PBHs is limited by
\begin{equation}
\Omega_{\rm PBH} \la \frac{4 \pi I_{\rm obs} M_{\rm PBH}}{L_{\rm ann} \Br (\gamma) \rho_c \int \delta (\vec{s}) ds}.
\end{equation}
To find the background $I_{\rm obs}$ the UCMH radiation competes with, we use the \emph{Fermi}-measured extragalactic background spectrum \citep{Abdo10} for $E \le 100\ \GeV$.  For annihilation into gamma rays or internal bremsstrahlung from charged particles, most of the power is within one log bin in energy of $m_{\rm DM} c^2$ \citep{Bell09}.  We therefore find $I_{\rm obs}$ by integrating the gamma-ray background as such.

We use an NFW density profile for the distribution of UCMHs in the Milky Way, $\delta(r) = \delta_s (r/r_s)^{-1} (1 + r/r_s)^{-2}$, with $\delta_s \approx 45000$ and $r_s = 27\ \kpc$ \citep{Stoehr03}.  The line of sight integral is not sensitive to the distribution of UCMHs in the inner Galaxy.  The integral is more like that for dark matter decay than diffuse annihilation, since the intensity is \emph{linearly} proportional to $n_{\rm PBH} (r)$.  We consider a sightline aimed directly away from the Galactic Center, where the uncertainty in the profile should have the least effect and the signal is smallest, for a conservative result.  

\begin{figure}
\centerline{\includegraphics[width=8cm]{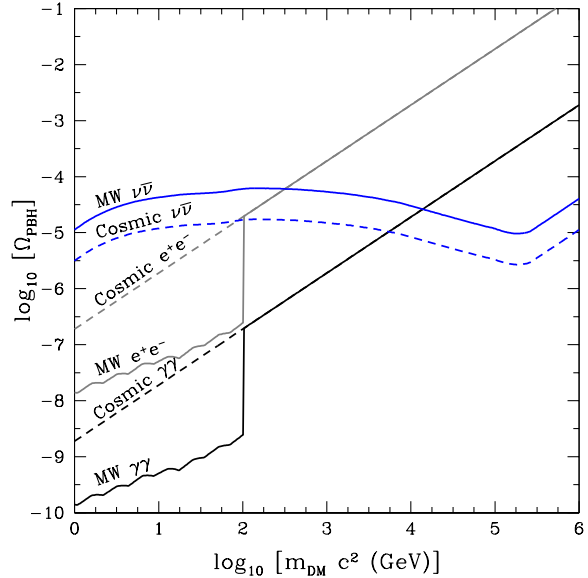}}
\figcaption[figure]{Upper bounds on the abundances of PBHs as a function of WIMP mass.  Bounds on annihilation into gamma rays (\emph{black}; $\Br (\gamma) = 1$) and electrons (\emph{grey}; $\Br (\gamma) = 0.01$) are shown, as well as neutrinos ($\Br (\nu) = 1$) (\emph{blue}).  Cosmic background limits are solid and Galactic limits are dashed.  Gamma-rays are the easiest final state to detect, while neutrinos are the hardest, and other Standard Model final states would give intermediate limits.\label{fig:BoundsWithMass}}
\end{figure}

The dependence of $\Omega_{\rm PBH}$ with WIMP mass is shown in Figure~\ref{fig:BoundsWithMass} for various final states.  The Galactic gamma-ray bounds in Figure~\ref{fig:BoundsWithMass} (\emph{solid}) include the cosmic background bounds above 100 GeV.  The annihilation luminosity falls as the WIMP mass increases (Eq.~\ref{eqn:LannPBH}).  Since the competing extragalactic background falls as $E^2 dN/dE \propto E^{-0.4}$ in the GeV to 100 GeV range, the allowed $\Omega_{\rm PBH}$ increases with WIMP mass.  The Galactic signal constrains $\Omega_{\rm PBH} \la 10^{-6}$ for a WIMP mass of 100 GeV even if $\Br (\gamma) = 0.01$, smaller than the cosmic bound (Eq.~\ref{eqn:CosmicGammaLimits}). 

With these abundances, we can calculate the lower limits on the mean distance to the nearest PBH, $\lambda_{\rm PBH} = [3 \pi / (4 \delta \mean{n_{\rm PBH}})]^{1/3}$, where the local $\delta = 89000$:
\begin{equation}
\lambda_{\rm PBH} \ga 220\ \pc\ m_{100}^{-1/3} \displaystyle \left(\frac{M_{\rm PBH}}{\Msun}\right)^{1/3} \left(\frac{\Br (\gamma)}{0.01}\right)^{1/3},
\end{equation}
for WIMP masses greater than 100 GeV.  This implies a $\gamma$-ray flux of
\begin{equation}
\Phi_{\gamma} \la 1.0 \times 10^{-9} \phFluxUnits m_{100}^{-4/3} \left(\frac{M_{\rm PBH}}{\Msun}\right)^{1/3} \left(\frac{\Br (\gamma)}{0.01}\right)^{1/3}.
\end{equation}

\section{Neutrino Background Constraints}
\label{sec:Neutrino}

Unlike gamma rays, neutrinos do not cascade down in energy as they travel through the Universe, although they redshift.  We integrate the atmospheric neutrino spectrum (or diffuse neutrino background limits above 100 TeV) from $m_{\rm DM} c^2 / e$ to $m_{\rm DM} c^2$ \citep{Gaisser02}, and require the neutrino flux from UCMHs be less than this; otherwise, they would have been detected \citep{Beacom07,Yuksel07}.  The measured data is reported in \citet{Ashie05}, \citet{Gonzalez06}, \citet{Achterberg07}, \citet{Hoshina08}, \citet{Abbasi09}, and \citet{DeYoung09}.

In Figure~\ref{fig:BoundsWithMass}, we show the Galactic (\emph{solid}) and cosmic (\emph{dashed}) bounds on PBHs from neutrinos.  The atmospheric neutrino background also steeply falls with energy ($E^2 dN/dE \propto E^{-1.3}$), so that the bound on $\Omega_{\rm PBH}$ is fairly energy-independent up to 100 TeV. 

\section{Conclusions}
\label{sec:Conclusion}
\begin{figure}
\centerline{\includegraphics[width=8cm]{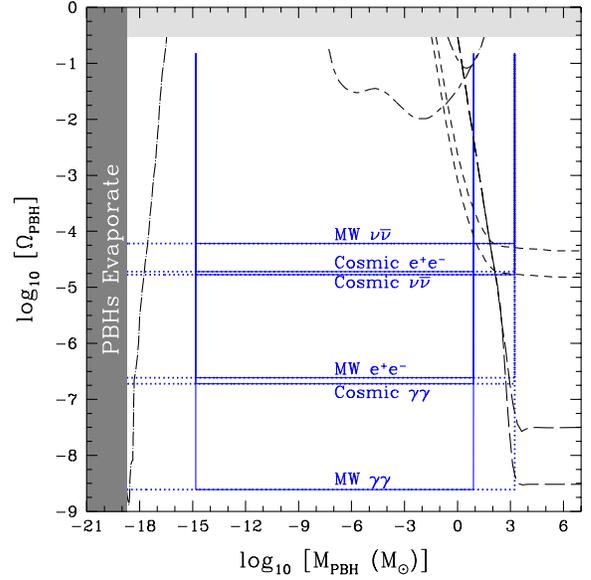}}
\figcaption[figure]{Upper bounds on the current abundance of PBHs of mass $M_{\rm PBH}$ for a WIMP mass of $100\ \GeV / c^2$ (\emph{left of the transition in Figure~\ref{fig:BoundsWithMass}}).  The limits in this work (blue; solid where baryonic opacity and non-radial accretion can conservatively be ignored, and dotted if baryonic opacity is small and accretion is radial) are the most powerful limits over a vast mass range.  See the mild caveats in the text.  Also shown are microlensing limits (\emph{short-dashed--long-dashed}), CMB limits from FIRAS (\emph{short-dashed}) and WMAP3 (\emph{long-dashed}), and evaporation radiation limits (\emph{dash-dotted}). See text for references. \label{fig:PBHFraction}}
\end{figure}

In Figure~\ref{fig:PBHFraction}, we show that even with the uncertainties in our estimates, our constraints are powerful for PBHs inside UCMHs.  PBHs of many masses are ruled out to very small abundances, for a wide range in WIMP masses.  At high PBH masses ($\ga 1\ \Msun$), CMB constraints become more powerful \citep{Ricotti08}; at very low masses, Hawking radiation limits are more powerful \citep[e.g.,][]{Carr09}.  Microlensing constraints are weaker, but make no assumptions about WIMPs \citep{Alcock01,Tisserand07}.  For $10^{-15}\ \Msun \la M_{\rm PBH} \la 10^{-9}\ \Msun$, ours are the \emph{only} constraints on PBHs aside from $\Omega_{\rm DM}$.  Our conclusion depends on the standard assumption that most dark matter is a self-annihilating thermal relic.  Our analysis does not apply if all of the dark matter is made of PBHs \citep[e.g.,][]{Frampton09}, because there will not be any WIMPs to annihilate.  For the smallest PBH and WIMP masses ($M_{\rm PBH} \la 10^{-15} m_{100}^{-3} \Msun$), radial accretion may not hold.  PBHs could remain important for other reasons \citep[e.g., as seeds of massive black holes; see][]{Mack07} which do not require them to have a substantial abundance.  The combination of our results with previous limits imply that PBHs either make up almost all of the dark matter, or almost none of it.  

Our results could be improved by detailed simulations of how the UCMH evolves as its inner regions annihilate.  Annihilation could lower the luminosity of the inner regions of the halo over time, weakening our bounds.  Steeper density profiles than $r^{-3/2}$ may increase the annihilation luminosity, but shorten the lifetime of WIMPs so that the halo annihilates away before $z = 0$.  We expect that some combination of constraints on gamma-ray and neutrino backgrounds, reionization history, and the CMB energy spectrum will generally require a small $\Omega_{\rm PBH}$.

Further improvements could come from more detailed studies of the annihilation products and their detectability.  When considering annihilation into charged particles, we considered only gamma rays from internal bremsstrahlung, but charged particles can themselves radiate, such as through Inverse Compton scattering \citep[e.g.,][]{Cirelli09}.  We are confident the limits for WIMP masses above 100 GeV can be improved by new observations.  Finally, \citet{Ricotti09} suggest that some UCMHs may exist without PBHs, formed from weaker initial perturbations in the early Universe.  Limits on dark matter annihilation in these UCMHs may strongly constrain these weaker perturbations.  

\acknowledgments
We thank A. Gould, J. Rich, M. Ricotti, K. Stanek, G. Steigman, and T. Thompson for helpful discussions.  This work was supported by NSF CAREER Grant PHY-0547102 to J.F.B. and an Alfred P. Sloan Fellowship to T. Thompson.

\end{document}